# Near-field Probing of Optical Superchirality for Enhanced Bio-detection

*Victor Tabouillot[1]\*, Rahul Kumar[1], Paula L. Lalaguna[1], Maryam Hajji[1], Rebecca Clarke[1], Affar Karimullah[1], Andrew R. Thomson[1], Andrew Sutherland[1], Nikolaj Gadegaard[2], Shun Hashiyada[3] and Malcolm Kadodwala[1]\**

[1] School of Chemistry, Joseph Black Building, University of Glasgow, Glasgow, G12 8QQ, UK

[2] School of Engineering, Rankine Building, University of Glasgow, Glasgow G12 8LT, U.K

[3] Department of Electrical, Electronic, and Communication Engineering, Chuo University, 1-13-27 Kasuga, Bunkyo-Ku, Tokyo 112-8551, Japan.

Malcolm.kadowala@glasgow.ac.uk, 2604448t@student.gla.ac.uk

ABSTRACT Nanophotonic platforms in theory uniquely enable < femtomoles of chiral biological and pharmaceutical molecules to be detected, through the highly localised changes in the chiral asymmetries of the near-fields that they induce. However, current chiral nanophotonic based strategies are intrinsically limited because they rely on far-field optical measurements that are sensitive to a much larger near-field volume, than that influenced by the chiral molecules. Consequently, they depend on detecting small changes in far-field optical response restricting detection sensitivities. Here we exploit an intriguing phenomenon, plasmonic circularly polarised luminescence (PCPL), which is an incisive local probe of near-field chirality. This allows chiral detection of monolayer quantities of a *de novo* designed peptide, which is not achieved with a far-field response. Our work demonstrates that by leveraging the capabilities of nanophotonic platforms with the near-field sensitivity of PCPL, optimal biomolecular detection performance can be achieved, opening new avenues for nanometrology.

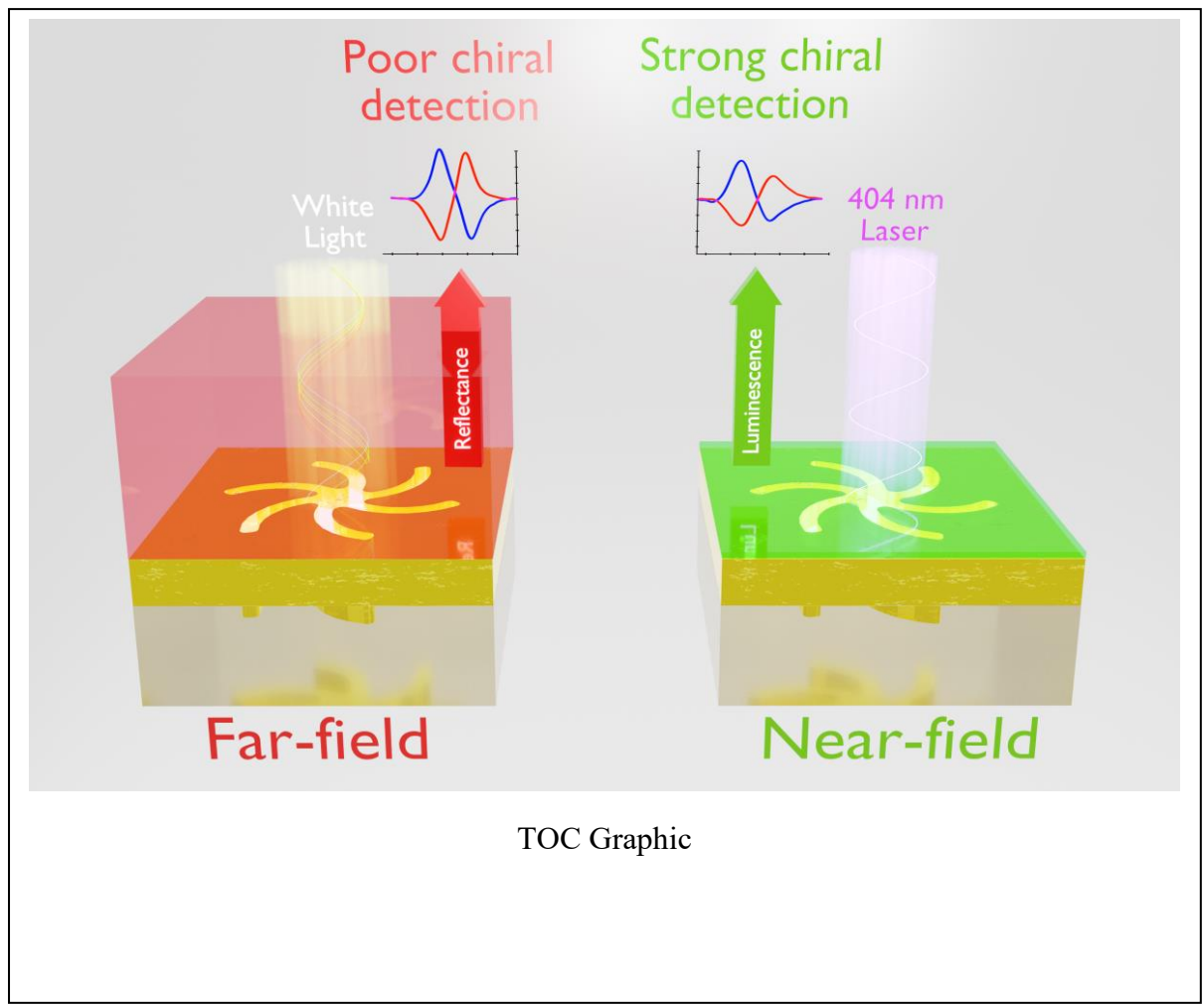

TOC Graphic

## Introduction

Chiral pharmaceutical and biological molecules are rapidly detected and characterised using chirally sensitive spectroscopic techniques based on the differential absorption and scattering of circularly polarised light (CPL). Potentially, chirally sensitive (chiroptical) spectroscopy could have broad impact, however, the inherent weakness of the dichroic interactions intrinsically limits applications. The sensitivity of chiral spectroscopic measurements can be vastly enhanced by leveraging the capabilities of nanophotonics to enhance chiral light matter interaction and thus make the chiral responses larger. The chiral light-matter interactions can be amplified using nanophotonics because near-fields with both enhanced intensities and chiral asymmetries (a property sometimes referred to as superchirality) can be created through the scattering of light from nanostructures. Using pairs of enantiomorphic plasmonic chiral nanostructures, near-fields of opposite symmetry can be produced, which interact asymmetrically with chiral media. This asymmetry causes differential changes in both the intensities and chiral asymmetries of the near-fields in the vicinity of the chiral media. These local changes in near-field properties are then detected through classical light scattering / absorption far-field measurements. The sensitivity of these types of measurements, like all plasmonic based sensing strategies, is limited by the volume occupied by the molecules of interest (which are typically monolayers adsorbed on to the surface of nanostructures) relative to the spatial extent of the near-fields. Consequently, plasmonic based sensors are less sensitive to relatively small molecules which occupy a smaller fraction of the near-field environment. This intrinsically limits the sensitivity and applicability of plasmonic based sensing techniques.

We report a novel phenomenon which enhances the sensitivity of plasmonic based chiral detection. Specifically, we demonstrate that plasmonic circularly polarised luminescence (PCPL) is far more sensitive to local changes in the optical chirality of the near-field than far-field light scattering-based measurements. Thus, a small chiral molecule can be detected, which is otherwise undetectable with traditional light scattering measurements. We attribute the enhanced sensitivity of PCPL to the fact that the signal is dependent on the local electromagnetic (EM) environment in the vicinity of the surface, the region occupied by adsorbed chiral molecules, rather than the significantly larger volume of the overall near-field from which the far-field response is derived. The ability of PCPL to provide greater sensitivities widens the potential application of chiral metamaterials for bio / enantiomeric detection.

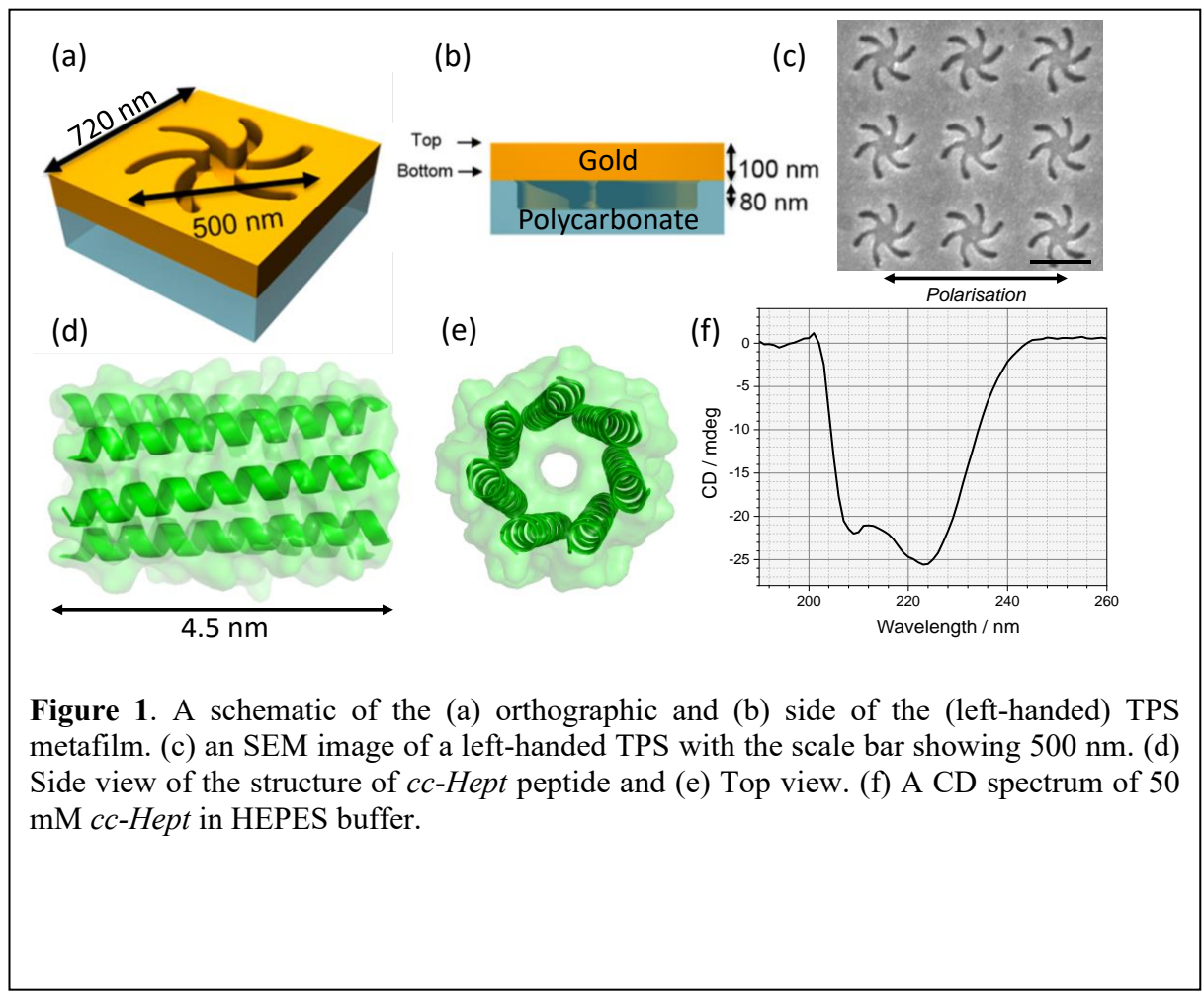

**Figure 1**. A schematic of the (a) orthographic and (b) side of the (left-handed) TPS metafilm. (c) an SEM image of a left-handed TPS with the scale bar showing 500 nm. (d) Side view of the structure of *cc-Hept* peptide and (e) Top view. (f) A CD spectrum of 50 mM *cc-Hept* in HEPES buffer.

# Background

***Metafilm samples***: The samples used is this study are made out of a ~100 nm thick gold metafilm deposited on a nanostructured polycarbonate template[1]. The polycarbonate platform consisted of either left-handed (LH) or right-handed (RH) "shuriken" shaped indentations, **Figure 1 (a)–(c)**, that possessed six-fold rotational symmetry and were arranged in a square lattice. These substrates are referred to as "template plasmonic substrates" (TPS) for brevity. The nanoscale indentations in the surface polycarbonate substrate have a depth of ~80 nm, are 500 nm in diameter from arm-to-arm, and have a pitch of 720 nm. A detailed discussion of the chiral and optical properties of these substrates can be found elsewhere[2].

***The de Novo Peptide***: As a well-defined, relatively small model biological molecule, we used a synthetic de novo designed heptameric $\alpha$-helical barrel assembly. Seven identical peptide chains self-assembled into a higher order $\alpha$-helical coiled coil. The resultant barrel structure is stable to chemical and thermal denaturation. To enable the peptide to be immobilised with a well-defined geometry on the surface, a thiol PEG linker is incorporated into the structure, **Figure 1 (d)–(e)**. For brevity, this coiled coil heptamer molecule will be subsequently referred to as *cc-Hept*. The CD spectrum of *cc-Hept* in buffer, **Figure 1 (f),** is qualitatively similar to that reported for a related $\alpha$-helical heptamer[3]. The immobilised peptide will have an approximate height above the surface of 4.5 nm.

***Basis of PCPL:*** It has been known for over 50 years that illuminating gold films with UV light results in photoluminescence which spans the visible to near-IR region[4]. This light emission is due to direct radiative recombination of electrons near the Fermi level with holes in the *d*-band. Early work also demonstrated that the amount of luminescence was enhanced by local fields generated by plasmonic excitation.[5] The basis of the PCPL strategy is that the polarisation properties of the luminescence is governed by the chiral asymmetry of the EM near-fields in the vicinity of the gold surface. The veracity of this assumption is provided by work which has demonstrated that achiral dye molecules embedded in a matrix surrounding a chiral plasmonic structure emitted CPL, and that the degree of circular polarisation, is correlated to the level of chiral asymmetry of the near-fields[6]. The chiral asymmetries of the near-fields in these previous studies, and others concerned with chiral nanophotonics, have been parameterised using an optical chirality (*C*) factor. The time average value of *C* is defined by:

$$C = -\frac{\omega}{2} Im(D^* \cdot B) \qquad (1)$$

Where $D = \epsilon(\omega)E$ and $B = \mu(\omega)H$ with $\epsilon = \epsilon' + i\epsilon''$ and $\mu = \mu' + i\mu''$ are the complex electric permittivity and magnetic permeability respectively, *E* and *H* are complex time harmonic electric and magnetic fields, and ω the angular frequency. The *C* factor is a conserved quantity of an EM field[7], and provides a convenient quantity for parameterising the chiral asymmetries of EM fields[8]. In more recent work,[9] it has been reported that optical chirality flux (*F*), which by analogy with Poynting's Theorem is identified in a conservation law for *C*, is proportional to the degree of circular polarisation. *F* is defined as:

$$\boldsymbol{F} = \frac{1}{4}[\boldsymbol{E} \times (\nabla \times \boldsymbol{H}^*) - \boldsymbol{H}^* \times (\nabla \times \boldsymbol{E})] \qquad (2)$$

In subsequent discussions of the PCPL, the observed levels of asymmetries are rationalised in terms of $F_z$, the component of the optical chirality flux parallel to the propagation of the incident beam.

***Optical measurements***: All optical measurements are collected from the TPSs using a reflection geometry. Spectra were obtained from enantiomorphic pairs of TPS while they are immersed in deionized water. Measurements were made on substrates which were either unfunctionalised or functionalised with *cc-Hept*. To provide achiral reference data sets for comparison we have also collected data from salt solutions.

***Far-field measurements***: The far-field chiroptical response was obtained using optical rotatory dispersion (ORD) which monitors the level of optical rotation (θ) as a function of wavelength. The ellipticity (ϕ) of the far-field reflected beam could not be directly determined due to limitation of the used setup. However, by using the *Kramers–Kronig* (*KK*) relationship ellipticity spectra can be derived from the ORD data.

***Near-field measurements***: Near-field measurements were collected from the same substrates as those used to obtain the far-field data. In contrast to the far-field measurements both θ and ϕ spectra are collected. The ϕ spectra are derived from measurements of the degree of circular polarisation of the luminescence.

***Parameterising spectral asymmetry***: Central to this study, is the potential for differential changes in chiroptical spectra, both light scattered (far-field) and luminescence (near-field) from LH and RH TPS induced by the presence of adsorbed chiral media. A factor, *A*, derived from the relative changes in the peak-to-peak height of resonances is used to parameterise the asymmetries in both optical rotation and ellipticity spectra. Derivatisation of the peak-to-peak heights and how these are used to obtain *A* is illustrated in **Figure 3** and **4**. This approach has been used to parameterise spectral asymmetries previously[10].

# Results

Far-field ORD data collected from *cc-Hept* functionalised and achiral salts solutions for both LH and RH substrates are shown in **Figure 2**, along with ellipticity spectra derived from the

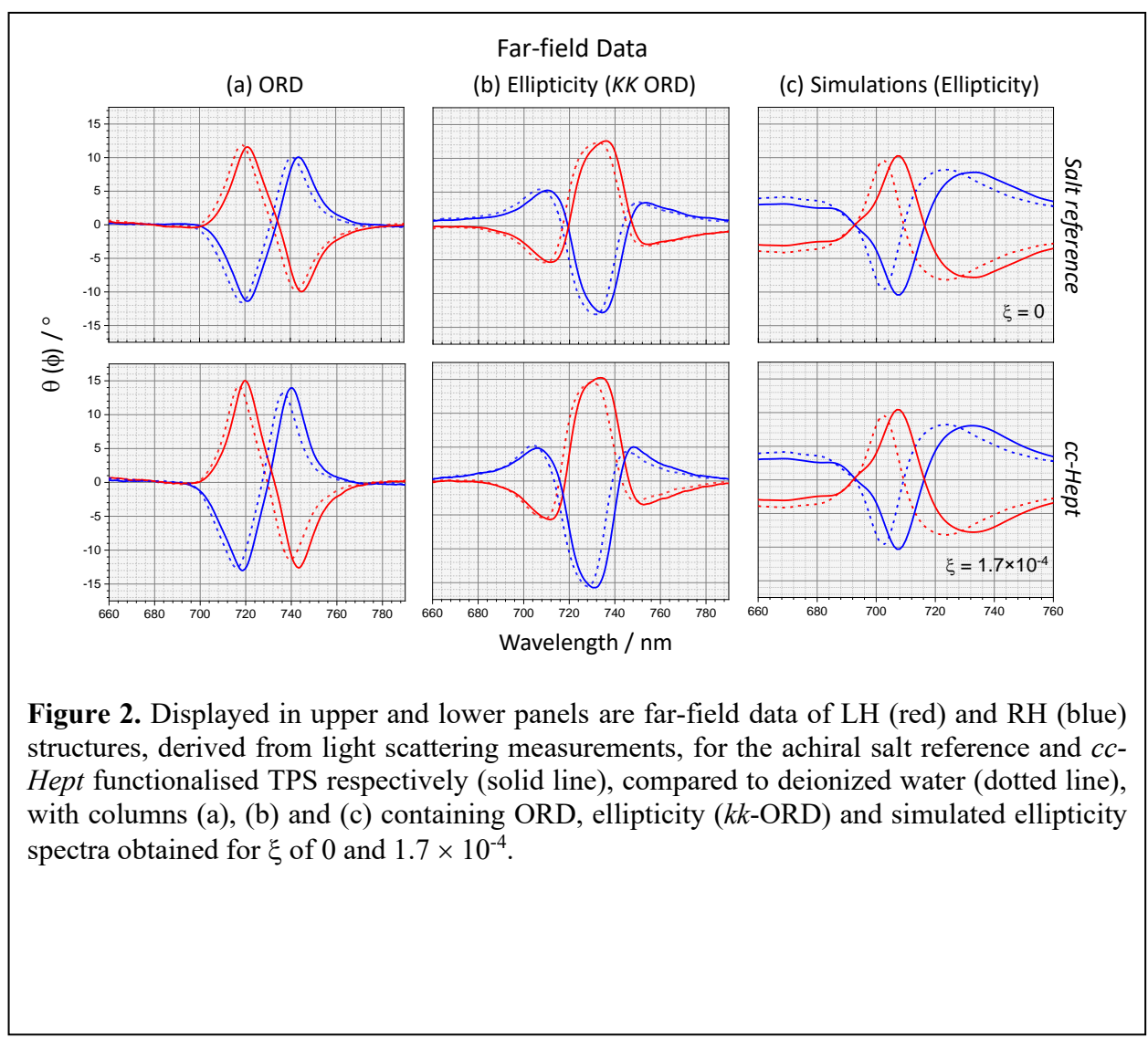

**Figure 2.** Displayed in upper and lower panels are far-field data of LH (red) and RH (blue) structures, derived from light scattering measurements, for the achiral salt reference and *cc-Hept* functionalised TPS respectively (solid line), compared to deionized water (dotted line), with columns (a), (b) and (c) containing ORD, ellipticity (*kk*-ORD) and simulated ellipticity spectra obtained for $\xi$ of 0 and $1.7 \times 10^{-4}$.

*KK* of the ORD. As expected, the adsorption of the *cc-Hept* causes red shifts in the positions of the resonances. There is no asymmetry observed for either the achiral salt solutions (as expected) or the *cc-Hept-* functionalised substrates, **Figure 5 (c)**.

The luminescence derived ϕ and θ spectra for both the achiral salt solutions and *cc-Hept* functionalised substrates are shown in **Figure 3** and **4**, with asymmetry factors given in **Figure**

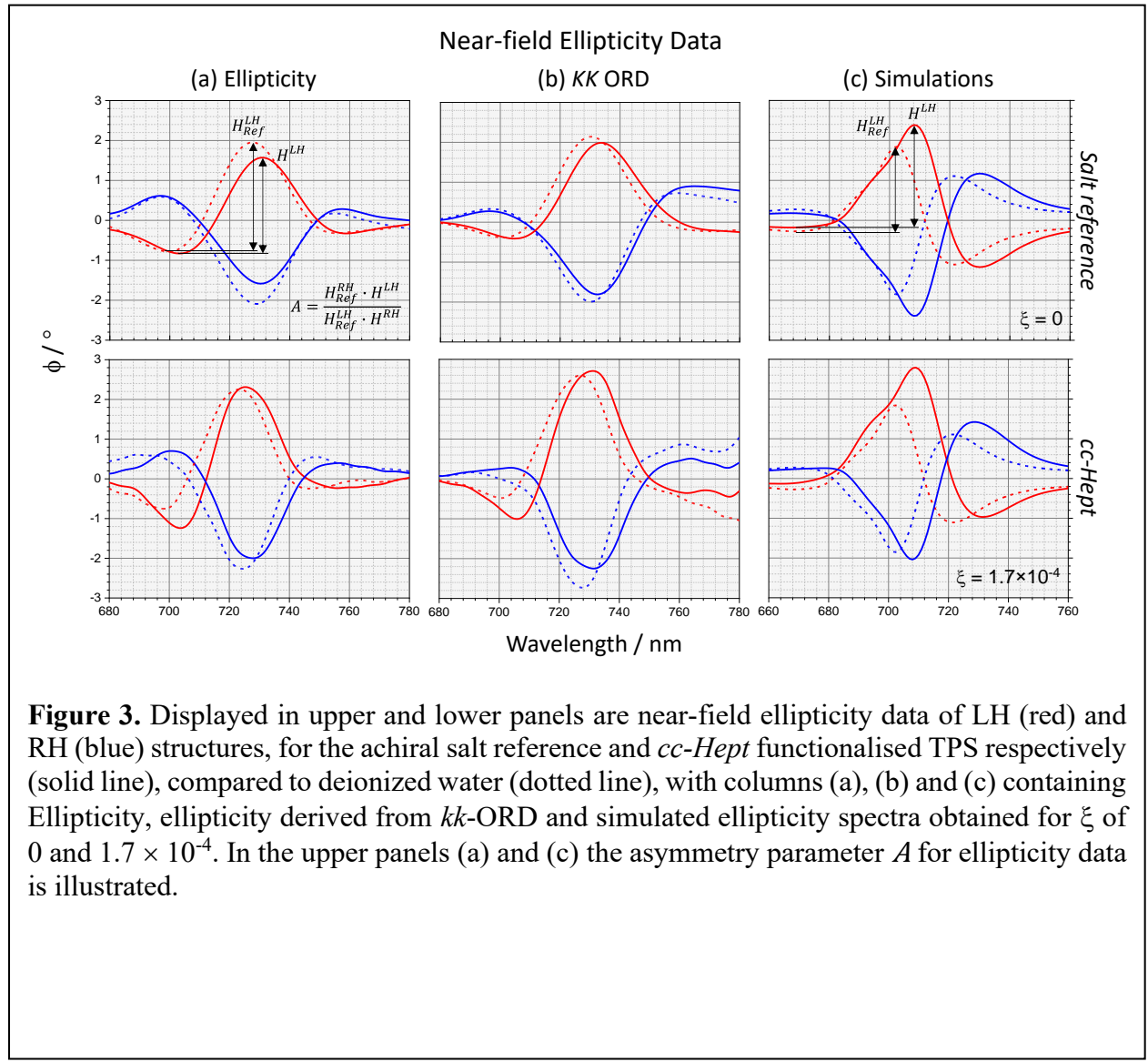

**Figure 3.** Displayed in upper and lower panels are near-field ellipticity data of LH (red) and RH (blue) structures, for the achiral salt reference and *cc-Hept* functionalised TPS respectively (solid line), compared to deionized water (dotted line), with columns (a), (b) and (c) containing Ellipticity, ellipticity derived from *kk*-ORD and simulated ellipticity spectra obtained for ξ of 0 and $1.7 \times 10^{-4}$. In the upper panels (a) and (c) the asymmetry parameter *A* for ellipticity data is illustrated.

**5 (c).** The similarity between the experimental and the corresponding *KK* data, demonstrates the robustness of the measurements. The *cc-Hept* data shows significant asymmetries, while no discernible asymmetries are observed for the achiral salt solutions data.

To validate the veracity of the light scattering and luminescence measurements we have performed EM numerical simulations. The modelling uses an idealised shuriken structure, and the *cc-Hept* layer is mimicked by a 10 nm thick isotropic chiral layer. The EM simulations are based on the implementation of the following constitutive chiral relationships: [9,24]

$$\boldsymbol{D} = \varepsilon_o \varepsilon_r \boldsymbol{E} + i\xi^T \boldsymbol{B} \tag{3}$$

$$\boldsymbol{H} = \boldsymbol{B}/\mu_0 \mu_r + i\xi^T \boldsymbol{E} \tag{4}$$

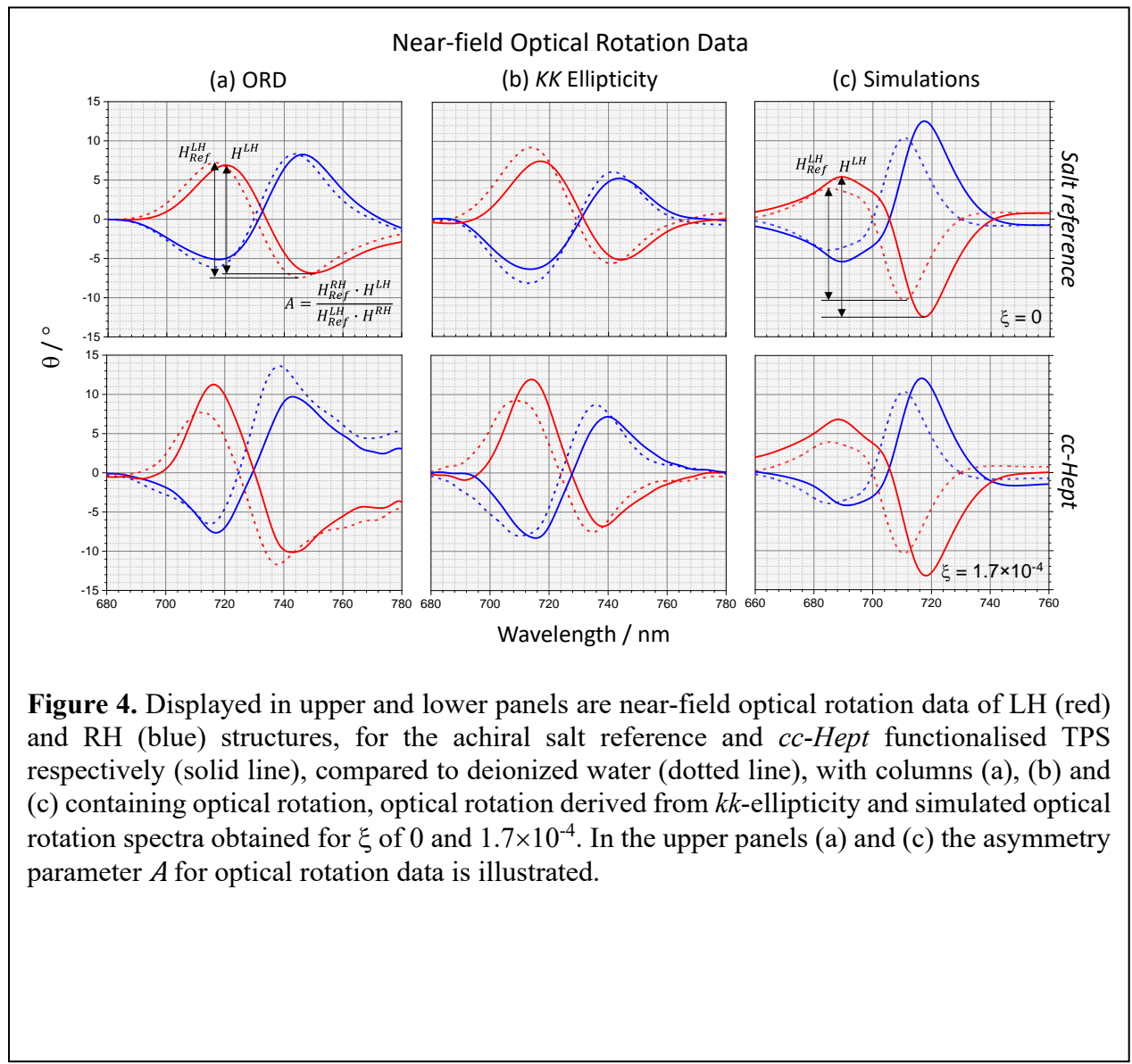

**Figure 4.** Displayed in upper and lower panels are near-field optical rotation data of LH (red) and RH (blue) structures, for the achiral salt reference and *cc-Hept* functionalised TPS respectively (solid line), compared to deionized water (dotted line), with columns (a), (b) and (c) containing optical rotation, optical rotation derived from *kk*-ellipticity and simulated optical rotation spectra obtained for ξ of 0 and $1.7 \times 10^{-4}$. In the upper panels (a) and (c) the asymmetry parameter *A* for optical rotation data is illustrated.

Here, $\varepsilon_o$ is the permittivity of free space, $\varepsilon_r$ is the relative permittivity, $\mu_0$ is the permeability of free space, $\mu_r$ is the relative permeability, $\boldsymbol{E}$ is the complex electric field, $\boldsymbol{B}$ is the complex magnetic flux density, $\boldsymbol{H}$ is the magnetic field and $\boldsymbol{D}$ is the electric displacement field. The chiral asymmetry parameter $\xi^T$ is a second rank tensor describing the chiral property of a medium, and is therefore zero for achiral materials. For an isotropic chiral layer $\xi_{xx} = \xi_{yy} =$

$\xi_{zz} \neq 0$, while all other elements are considered zero[2, 11] In the current simulations, we have used a $\xi_{xx} = \xi_{yy} = \xi_{zz} = 1.7 \times 10^{-4}$ which is consistent with values used in numerical simulations in previous work to replicate the effects of protein layers[10-12]. The numerical simulations have been used to calculate volume integral values of $F_z$ in two regions of space above the surface of the TPS, **Figure 5 (a)–(b)**, within the 10 nm thick chiral layer and the entire volume above the surface of the TPS including the chiral layer. The $F_z$ values were normalised by that for left CPL, producing a number which is equivalent to the degree of circular polarisation, which can then be converted into $\phi$ spectra. Afterwards, equivalent $\theta$ spectra can be obtained from *KK* transformation. These simulated spectra replicate the experimental luminescence and lights scattering experiments. This is illustrated by the good agreement between the *A* values obtained from the simulated spectra and those derived from experiment, **Figure 5 (c)**.

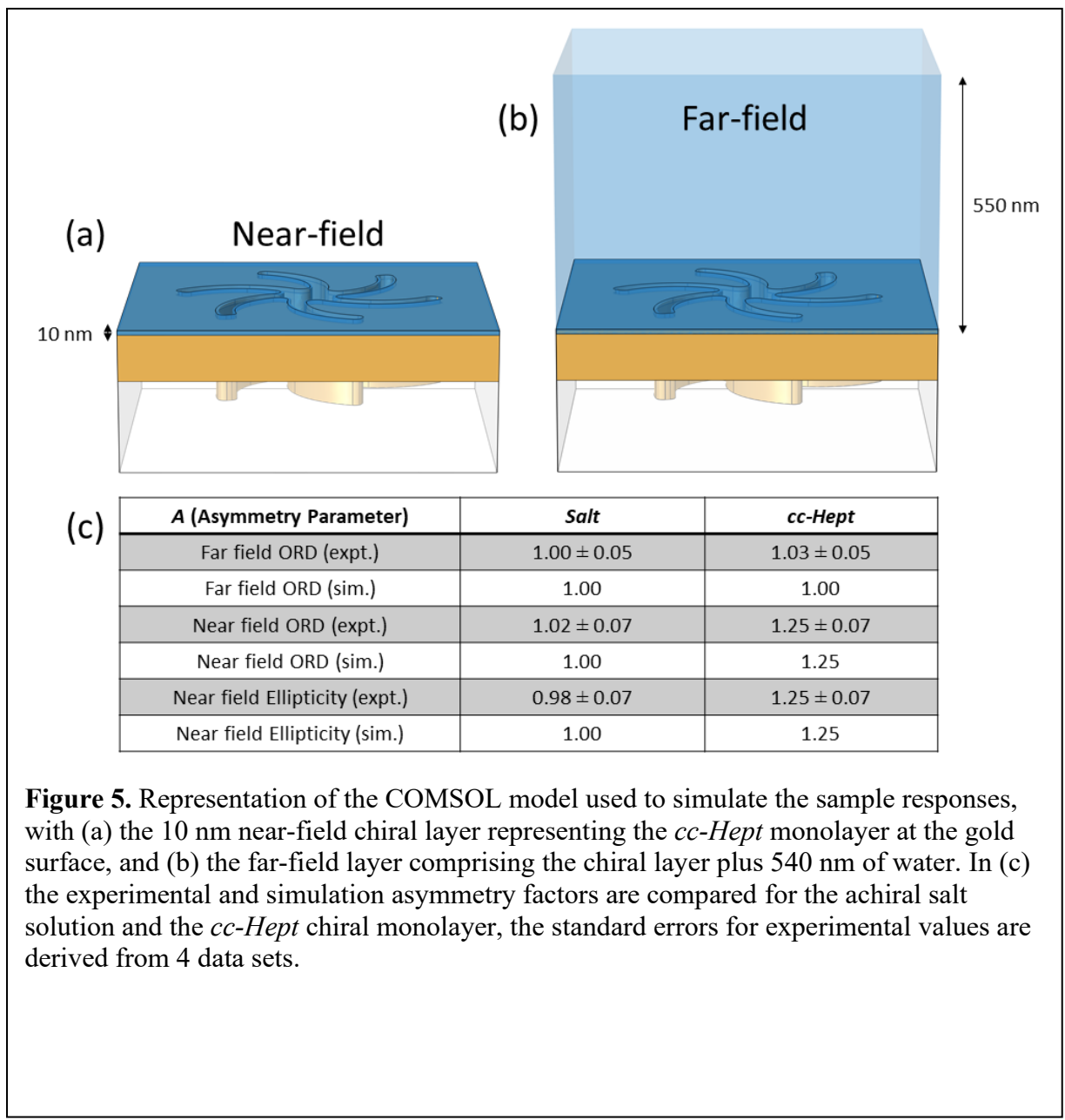

**Figure 5.** Representation of the COMSOL model used to simulate the sample responses, with (a) the 10 nm near-field chiral layer representing the *cc-Hept* monolayer at the gold surface, and (b) the far-field layer comprising the chiral layer plus 540 nm of water. In (c) the experimental and simulation asymmetry factors are compared for the achiral salt solution and the *cc-Hept* chiral monolayer, the standard errors for experimental values are derived from 4 data sets.

## Discussion

We have shown unambiguously that chiroptical measurements based on luminescence from plasmonic meta-films can detect a chiral response from a *de novo* designed peptide which is undetectable with classical light scattering based measurements. The enhanced sensitivity of luminescence measurements is attributed to the fact that they are more sensitive to the surface region, which is occupied by the adsorbed chiral molecule, than the far-field light scattering data. This hypothesis is supported by numerical simulations which replicate the experimentally observed behaviour. The work we report clearly demonstrates that bio / enantiomeric detection sensitivity can be enhanced by using luminescence from the Au plasmonic metafilm as a local probe of the EM environment occupied by the adsorbed chiral molecules. This new paradigm for exploiting (chiral) nanophotonic platforms for chiroptical measurements has potential applications in nanometrology and point of care diagnostics.

## Methods
### Sample nanofabrication

The polycarbonate slides which had the shuriken nano-indentations were manufactured using an injection moulding machine (ENGEL) was described in detail elsewhere[13]. The shuriken indentation on the substrate had a depth of ~80 nm and a length of 500 nm from arm to arm. After fabrication, the slide was cleaned with IPA and dried with $N_2$ gas. The cleaned slide was metal evaporated using a Plassys MEB-400s to a gold thickness of 100 nm and then cleaned in an oxygen plasma asher (see *Supplementary Information*).

### *cc-Hept* functionalisation

The *cc-Hept* monolayer was deposited on the gold substrate after conducting all the measurements in deionised water. Oxygen plasma treatment was conducted for 1 minute at 80 W before placing the sample in a 0.1 mM solution of *cc-Hept* diluted in HBS (HEPES buffered saline, 10 mM HEPES and 150 mM NaCl in water at pH 7.2). After 24 hours of incubation, the sample was rinsed using HBS to flush out the unbound molecules.

### Far-field Optical rotatory dispersion (ORD) measurements

ORD measurements of our samples were performed with custom-made polarisation microscope described previously. The instrument consists of a tungsten halogen lamp (Thorlabs) light source that propagates through a collimating lens, Glan-Thomson polariser (Thorlabs), 50:50 beamsplitter and a focused beam 10× objective lens (Olympus). The reflected beam propagates back from the second Glan-Thompson polarisers (Thorlabs) and is captured by the spectrometer (Ocean Optics USB400). The sample was positioned and focused using a camera (Thorlabs DCC1645C) after the analyser, allowing the polarised beam to hit the nanostructures at the orientation shown in **Figure 1 (c)**. The Stokes methods are using to record the ORD spectra at four analyser angles (0°, ±45° and 90°) concerning the incident polarisation for enantiomorphic pairs of TPSs (see *Supplementary Information*).

### Photoluminescence setup

Photoluminescence measurements were carried out using a home-built microscope from Thorlabs (ESI). A 404 nm laser diode, with a 180-mA fixed current and maximum optical power output of ~17 mW, was used as the excitation source. Two linear polarisers at the input after the laser diode were mounted to adjust the input power. The first polariser was used to modify the input power and the second polarised to define the input polarisation. The beam was focused using a 10× (NA=0.3) objective and the photoluminescence signal was collected with the same objective in a reflection geometry. The position of the sample and alignment of the laser were monitored using an optical camera. The sample was positioned as such that the polarised beam was at the same orientation as for the far-field measurements, as shown in **Figure 1 (c)**. A CCD camera mounted on the top of the configuration was used to capture the signals (see *Supplementary Information*).

**Numerical electromagnetic modelling**

A commercial finite element package (COMSOL V5.6 Multiphysics software, Wave Optics module) has been used to simulate the intensity and chiral asymmetry of EM fields produced across the sample. Periodic boundary conditions have been imposed on the sides of the meta-film (i.e., equivalent to simulating a meta-film array). Reflections were minimised using a perfectly matched layer (PML) above and below the input and output ports.

Linearly polarised EM waves were performed at normal incidence onto the meta-film. The finite-element method has been used to solve Maxwell's equations over a distinctive geometry which allows the optical chirality and E-field intensity to be measured in COMSOL. A 10 nm continuous dielectric domain, with a refractive index of 1.44, has been defined at the surface of the sample to simulate the *cc-Hept* monolayer. Above this layer, a 540 nm domain containing water ($n = 1.331$) has been defined on top of the chiral layer, allowing the calculation of the far-field volume average of $F_z$ (see *Supplementary Information*).

ACKNOWLEDGMENT

Technical support from the James Watt Nanofabrication Centre (JWNC) and VT acknowledges the EPSRC for the award of a studentship.